\documentclass[preprint]{vgtc}               % preprint
\graphicspath{{figures/}{pictures/}{images/}{./}} % where to search for the images

\usepackage{times}                     % we use Times as the main font
\usepackage{tabu}                      % only used for the table example
\usepackage{booktabs}                  % only used for the table example
\usepackage{colortbl}
\usepackage{listings}                  % used to typeset agent prompts in the appendix
\usepackage{lipsum}                    % used to generate placeholder text
\usepackage{mwe}                       % used to generate placeholder figures

\usepackage{mathptmx}                  % use matching math font

\onlineid{1421}

\vgtccategory{Research}

\vgtcinsertpkg

\title{A Design Study on Voice-based Interaction for Immersive Network Visualization and Analysis}

\author{Sam Yu-Te Lee\thanks{e-mail: ytlee@ucdavis.edu}\\ %
        \scriptsize University of California, Davis %
\and Hsin-Ai Chen\thanks{e-mail: xacchen@ucdavis.edu}\\ %
     \scriptsize University of California, Davis %
\and Sarah Yuniar\thanks{e-mail: sayuniar@ucdavis.edu}\\ %
     \scriptsize University of California, Davis %
\and David Bauer\thanks{e-mail: davbauer@ucdavis.edu}\\ %
     \scriptsize University of California, Davis %
\and Kwan-Liu Ma\thanks{e-mail: klma@ucdavis.edu}\\ %
     \scriptsize University of California, Davis %
}
\abstract{
       Visual network analysis leverages network visualization authoring techniques to facilitate sensemaking, serendipitous discovery, and hypothesis verification on network data. However, transferring the same paradigm to immersive environments is non-trivial due to insufficient UI affordance for authoring operations. Researchers have studied combining multiple modalities for interactions, but the high learning curve of such input systems limits their adoption by typical data analysts, let alone for network analytics. In this work, we investigate the advantages and limitations of voice as the primary input modality with a research-through-design (RtD) study, in which we design a system that supports voice-based interactions for immersive network visualization facilitated by Large Language Models (LLMs). Through a user study on social network data analysis with participants from social science and computer science backgrounds, we find that voice interactions can improve perceived usability relative to controller-based interaction and lower the cognitive effort of formulating commands, since users can express intent in natural language rather than compressing it into terse instructions. We discuss design implications for immersive visualizations, highlighting how usability limits adoption while simplified interactions and voice-based controls enhance fluidity and support complex, multi-parameter operations.
} % end of abstract


\begin{document}

\maketitle
\section{Introduction}
% Introduce the real-world scenarios of network analysis done by domain experts
Beyond calculating statistical measures of networks, visual investigation is fundamental to understanding network structure in domain-specific contexts and often surfaces new hypotheses~\cite{tommaso2021visual_network_analysis, markus2020visual_network_research}.
Network visualization authoring systems~\cite{bastian2009gephi, shannon2003cytoscape} let users construct network visualizations with different layouts and visual encodings, directly supporting this form of investigation.

% What immersive network analytics promise
Immersive network analytics promises to advance such analysis by leveraging the immense display space and enhanced spatial reasoning~\cite{kraus2021value_of_immersive_vis}.
% However, such an investigation on 2D interfaces is limited by the visual scalability of the authoring design as well as available screen space.
% Immersive network analysis has the potential to overcome this issue with access to an immense display space and enhanced spatial reasoning~\cite{kraus2021value_of_immersive_vis}.
% highlight the complexity of network operations
However, the interaction design developed for network authoring on 2D interfaces cannot be trivially translated to immersive environments, where UI affordances are limited. 
Network authoring encompasses a wide range of operations, such as selecting nodes or changing layouts~\cite{lee2006graphtaxonomy, saket2014group_level_graph_taxonomy}, each of which needs to be mapped to a user input signal. 
Naively binding operations to UI elements risks producing excessive selection menus in a typical WIMP (Windows, Icons, Menus, and Pointers) interface~\cite{Teyseyre2009overview3dsoftware}.

% Introduce current approaches on multi-modal input, and their limited usability, resulting in low adoption
Researchers have explored multimodal inputs for immersive analytics, such as touch, gaze, mid-air gestures, and speech~\cite{badam2017affordances_immersive_env}. 
Since different modalities suit different tasks, combining them is widely thought to let the strengths of one offset the limitations of another~\cite{srinivasan2018orko, saktheeswaran2020touch_and_speech}. Yet designing such combinations is not straightforward, since user preferences for modalities vary across tasks, and the wide range of operations in immersive network authoring makes the design space demanding~\cite{yijheng2017gesturegraph}. An interaction design that achieves fluidity~\cite{elmqvist2011fluid_interaction} for immersive network authoring, where users can express and execute authoring operations smoothly while keeping their attention on the data, has yet to be established.
% The extensive authoring operation space, coupled with the combinatorial complexity of modality integration, poses significant challenges for interaction design.

% Introduce our proposal of using speech as the primary input, and why we believe this will work
% While recent approaches seek to combine multiple input modalities to replace the traditional WIMP interface~\cite{saktheeswaran2020touch_and_speech, dhanoa2026heydashboard}, 
Rather than adding more modalities, we approach this challenge from a different angle: we propose a multimodal paradigm in which voice serves as the primary input, supplemented by minimal controller-based interaction.
Our rationale is two-fold. First, voice commands reduce the specification cost of network authoring. Queries such as ``color the female smokers in red and bring them closer'' or ``select the students with GPA above three and color them by grade'' bundle predicates and several coordinated parameters into a single utterance, which would require a sequence of operations on many UI elements in a traditional UI.
Second, voice input is increasingly embedded in LLM-based software (e.g., ChatGPT) and devices (e.g., smart glasses) to manage broad command spaces while improving usability, demonstrating that speech recognition is now robust enough to enhance usability in deployed products, not just research prototypes.

Still, the advantages and limitations of voice as the primary modality for immersive network authoring remain underexplored~\cite{badam2017affordances_immersive_env}. 
To address this gap, we present a research-through-design study in which we characterize the requirements of voice-based immersive network interaction, design and implement a voice-based multimodal system to meet them, and examine it through a qualitative user study.
The system builds on a taxonomy~\cite{lee2006graphtaxonomy} of network authoring and analysis operations for immersive environments, together with an LLM-based recognition pipeline that maps voice commands to operations. The recognized commands are then interpreted by our immersive rendering module and executed, enabling network authoring entirely through voice.

% Introduce our insights
Through a user study on social network analysis, we find that voice-primary interaction improves usability by reducing physical effort and lowering the cognitive cost of formulating commands. We then discuss how usability shapes VR adoption and how voice-primary interaction improves interaction fluidity. Our contributions are twofold: a voice-based multimodal system that instantiates a voice-primary design for immersive network authoring, serving as a research artifact for examining this underexplored interaction paradigm; and, through a qualitative study of its use, a set of design implications and reflections on how voice-primary interaction shapes usability, fluidity, and adoption in immersive analytics.

% Name our contribution
% Our contributions are:
% \begin{itemize}
%     \item A voice-based immersive network analytics system with empirical evaluation on its usability,
%     \item Insights on how voice-based interactions improve usability in immersive network analytics. 
% \end{itemize}
\section{Related Works}
Our work builds on research in immersive network analytics. We review how voice-based interaction can enable more intuitive and efficient exploration of complex networks while complementing traditional visual and manual inputs in immersive environments.

\subsection{Visual Exploration and Analysis of Networks}
Visual exploration and analysis of networks has been extensively studied by the visualization community~\cite{nobre2019sota_multivariate_networks, von2011sota_large_graph}, with representations such as matrices~\cite{Han2024Nuwa}, arcs~\cite{dang2016timearcs}, and biofabrics~\cite{fuchs2026biofabric}.
In practice, the most used visual representation is still the node-link diagram~\cite{shannon2003cytoscape} for its general availability in commercial tools. 
To visually explore and analyze networks, users would change layouts and encodings of the network to reveal interesting visual patterns~\cite{Han2024Nuwa}.     
To comprehensively understand the tasks associated with network exploration, researchers have proposed task taxonomies~\cite{lee2006graphtaxonomy, saket2014group_level_graph_taxonomy} that cover finding, coloring, and sizing nodes and edges, in addition to typical graph-related arithmetic such as degree and centrality calculation.
AlKadi et al.~\cite{alkadi2023barrier_to_network_exploration} conducted a study to understand barriers to visual network exploration and found that analysts have difficulty learning and interpreting the excessive number of authoring operations, calling for future tools to ease cognitive load. 
In our work, we investigate the effectiveness of using voice as the primary interaction for authoring networks in immersive environments.

\begin{figure*}[t]
 \centering % avoid the use of \begin{center}...\end{center} and use \centering instead (more compact)
 \includegraphics[width=\textwidth]{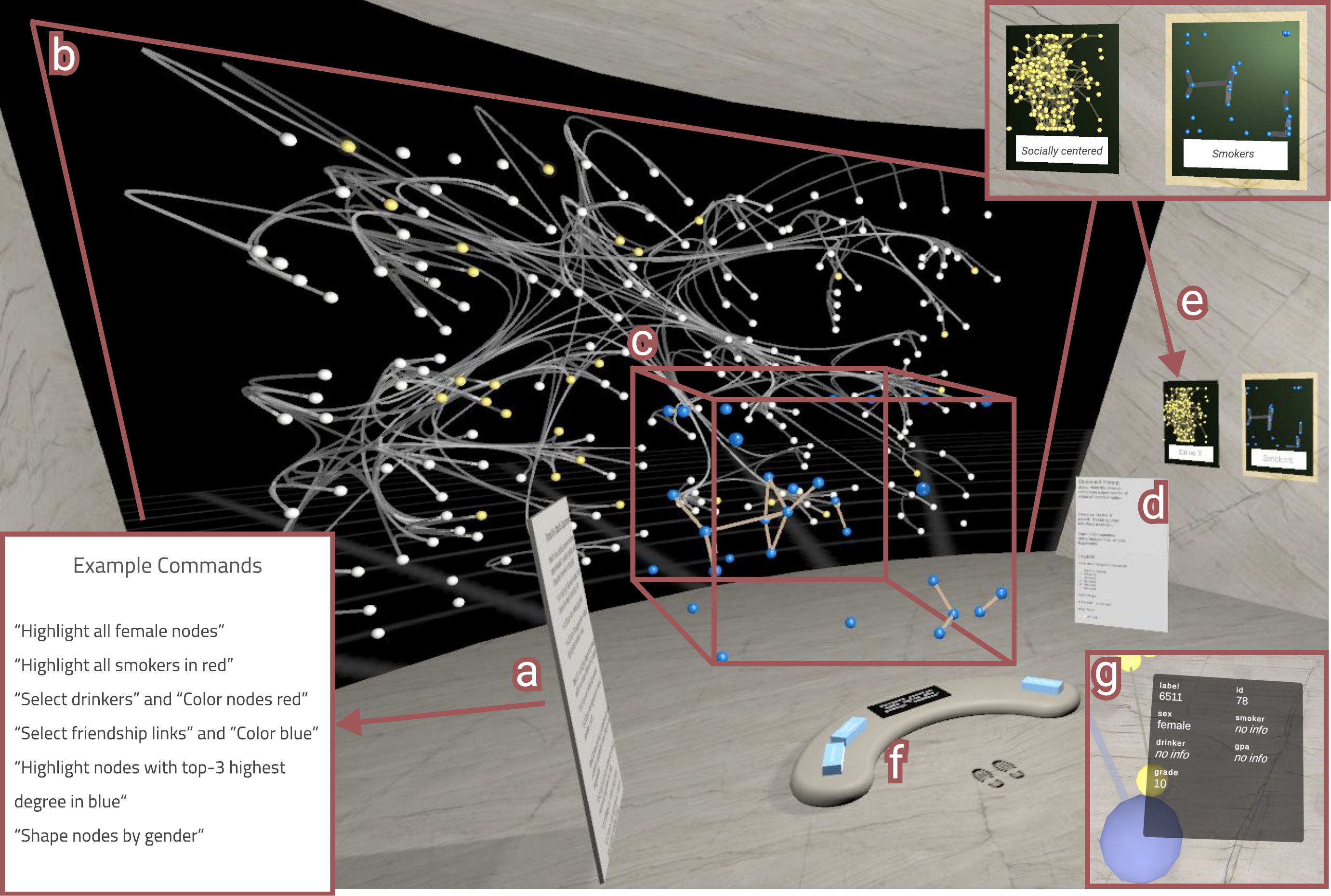}
 \caption{Overview of the system. (a) an example command panel illustrating supported commands (b) main graph showing all the nodes and links in the dataset (c) a subgraph created by the user that can be further authored on (d) text panel showing command history and legends (e) session gallery showing analysis history saved by the user (f) a dashboard with buttons for creating, saving, and deleting sessions, (g) an example tooltip when the user points to a node using the controller.}
 \label{fig:pov}
 \vspace*{-0.5cm}
\end{figure*}

\subsection{Immersive Network Visualization}
Immersive network visualization uses virtual reality devices to provide users with enhanced sensory perception and embodied interaction~\cite{Ens2021grand_challenges} to aid data understanding and analysis. 
Despite well-known limitations to 3D data visualizations~\cite{munzner2014visualization_analysis_design}, immersive network visualization promises to enhance spatial understanding~\cite{batch2020there_is_no_spoon, kotlarek2020mental_maps_immersive}, increase engagement and intuitiveness, better support complex and large networks, and better support collaborative analysis~\cite{joos2025visual_network_analysis_immersive_survey, kraus2021value_of_immersive_vis, cordeil2017cave_vs_hmd_network}.
However, appropriate designs for supporting visual exploration and analysis of networks in immersive environments remain underexplored. Joos et al.~\cite{Joos2022visual_comparison_network_vr} studied the appropriate visual representations for network comparison, and found that node-link diagrams are significantly better than matrix representations, which is contrary to comparable experiments in 2D. Kwon et al.~\cite{kwon2016layout_rendering_interaction} proposed advanced layout and edge routing techniques to improve visual clarity of node-link diagrams in immersive environments. Bauer et al.~\cite{bauer2024multi_layout_network} introduced a multi-layout design to promote more efficient use of space for hierarchical network exploration. These works collectively show the value and potential of immersive network visualization, but designing a system that supports a large number of interactions remains challenging~\cite{Ens2021grand_challenges}. We seek to improve the usability of immersive network visualization systems through voice-based interactions.

\subsection{Multi-modal Immersive Analytics}
Interaction is widely regarded as central to realizing the benefits of immersive analytics~\cite{marriott2018immersive_analytics, buschel2018interaction_for_ia, skarbez2019ia_research_agenda}. Because no single input modality suits every task, researchers have explored combining touch, gaze, mid-air gesture, and speech so that the strengths of one modality offset the limitations of another~\cite{bolt1980put_that_there, badam2017affordances_immersive_env, saktheeswaran2020touch_and_speech, yijheng2017gesturegraph}. However, combining modalities increases the learning curve, and the design complexity grows with the large operation space of network authoring, as user preferences for modalities also vary across tasks~\cite{saktheeswaran2020touch_and_speech}. Within these combinations, voice has typically been positioned as a complementary modality rather than the primary one.

We argue that voice is particularly well suited as the \emph{primary} modality for immersive network analytics for two reasons. First, its advantage is relative to the input each setting otherwise relies on: on 2D/2.5D screens, mouse and keyboard already offer strong usability, whereas in immersive environments the default alternatives (handheld controllers and hand tracking) are slow, imprecise, and error-prone for text entry and precise selection~\cite{dube2019text_entry_vr, grubert2020back_to_the_future}. In comparison, voice yields a relatively larger practical gain. Second, network analysis tasks are relational and multi-parameter~\cite{saktheeswaran2020touch_and_speech}: a single utterance can bundle predicates and coordinated parameters that would otherwise require a sequence of operations across many UI elements. Recent advances in speech recognition and language understanding now make such voice-primary interaction practical, which we investigate in this work.

% These are NLI works and not necessarily in immersive space
After speech recognition transcribes a spoken command into text, voice-primary interaction closely resembles a natural language interface (NLI), which is known as an intuitive, novice-friendly paradigm for visualization tasks~\cite{Shen2023NLIsurvey, narechania2021nl4dv}.
Similar to our motivation, Srinivasan et al.~\cite{srinivasan2018orko} proposed Orko, an NLI for 2D network authoring, and received positive feedback. Yet it is unknown if the same design applies to an immersive environment.
This line of work also surfaces challenges that persist in NL interfaces: resolving ambiguity in utterances~\cite{gao2015datatone}, improving the discoverability of commands~\cite{srinivasan2021snowy}, and grounding interpretation in realistic user utterances~\cite{srinivasan2021NL_utterances, song2025embodied_nli}.
Recent advances in large language models (LLMs) have substantially improved natural language understanding, renewing interest in NLIs~\cite{wen2025multi_modal_prompt, dhanoa2026heydashboard, dibia2023lida}. Combined with speech recognition, voice input can now be seamlessly integrated into LLM-based interactions, motivating exploration of voice interfaces for immersive systems. However, it remains unclear whether LLM-based voice interaction improves usability in immersive network authoring~\cite{badam2017affordances_immersive_env}, and we aim to address this gap.
\section{System Overview}
In this section, we introduce our system that uses voice as the primary input modality for immersive network visualization and analysis, including the network visualizations, system commands, and a multi-agent architecture for command classification. The system is open-sourced (\url{https://github.com/sarahayu/VR-Network-Visualization.git}) for reproduction.

\subsection{Main Graph and Subgraph View}

% The immersive system is implemented in Unity~\cite{unity}, a popular game engine for constructing video games with the capability for VR and AR features. 
%
% It comes with the builtin library XR Interaction Toolkit~\cite{InteractionToolkit} which bootstraps much of the VR implementation, such as detecting VR input and providing templates for ``interactables'', objects in the game scene that can be moved around by the user or pressed like buttons. 
The system starts with a three-walled room with a main graph in front, as shown in \autoref{fig:pov}-b (the subgraph generated through user interactions, labeled (c) in the figure, is not shown at startup). 
The room provides a sense of scale and helps ground the user in the immersive space, while the absence of the back wall allows the graph to appear unencumbered without feeling cramped.
The main graph uses a spherical layout with edge bundling that was shown to be an effective method for immersive network visualizations~\cite{kwon2016layout_rendering_interaction}. 
The main graph serves as an overview and is uneditable, i.e., node attributes such as position, color, and size cannot be changed by the user. All network exploration and analysis is performed on the dynamically generated subgraph view. 

The subgraph view is at the center of the room and displays a user-specified subgraph using a 3D force-directed layout without edge bundling.
Users can issue authoring commands through voice (e.g., change node and link color) as specified in~\autoref{sec:commands}, or use controllers to select individual nodes manually. 
Nodes included in the subgraph are simultaneously highlighted in yellow in the main graph, allowing users to maintain awareness of their location within the overall network context. The main graph can be hidden to prevent overlapping with the subgraph from a certain point of view.
Placing the subgraph at the center allows users to walk around the subgraph for multi-angle observation and address node occlusion. 
Pointing at a node using the controller highlights its connected edges and triggers a tooltip describing its data attributes (\autoref{fig:pov}-g).
The tooltip scales with distance, appearing larger at farther distances from the user and appearing smaller at closer distances to maintain readability, and renders above all elements of the scene to prevent occlusion.

%
% The tooltip also renders above all other elements of the scene to prevent occlusion from other objects and contribute to enhanced readability.

% After a session is created and a voice query is made, a second node-link graph appears in the empty space between the dashboard and the main graph, referred to as the "subgraph". 
%
% Each session has a subgraph associated with it, and only one subgraph may be visible in front of the user at any time (the main graph remains visible at all times).

% The subgraph is visually similar to the main graph, with the exception being the links are straight, not bundled, and that it is aligned spatially via a force-directed system.
%
% The subgraph is also editable and the user may move nodes of the subgraph around if they wish by pointing the controller at a node, holding down the grip button, and moving the controller around.

% In a similar fashion, the user may also select nodes of the subgraph for editing, but instead of holding down the grip button, the user may simply press and quickly release the grip button (similar to clicking a mouse button) while pointing at a node. 
% %
% Users can deselect nodes by repeating the same action, or they can deselect all nodes by clicking an empty space. 

\begin{table*}[t]
\centering
\begin{tabular}{p{3cm} p{8cm} p{5cm}}
\toprule
\textbf{Command} & \textbf{Description} & \textbf{Examples} \\
\midrule

Selection 
& Selecting nodes or links with data attributes or visual attributes. Subsequent commands default to the selected subset. 
& Select female nodes; Select red nodes. \\
\arrayrulecolor{gray}
\specialrule{0.2pt}{0pt}{0pt}
Set Color
& Set node or link color, or encode an attribute with color.
& Color nodes by gender. \\
\specialrule{0.2pt}{0pt}{0pt}
Set Node Shape 
& Encode node shape with a categorical attribute. 
& Encode gender with shapes. \\
\specialrule{0.2pt}{0pt}{0pt}
Change Node Position 
& Adjust node positions in the scene (e.g., bringing nodes closer to the user). 
& Bring high-GPA nodes closer to me. \\

\specialrule{0.2pt}{0pt}{0pt}Statistical Query 
& Perform statistical and arithmetic operations.
& Encode node degree with colors; What's the degree of this node? \\
\arrayrulecolor{black}

\bottomrule
\end{tabular}
\caption{Supported voice command categories in our system, including their functionality and example utterances used to perform network exploration and analysis tasks.}
\label{tab:commands}
\end{table*}
\subsection{System commands}\label{sec:commands}

The commands supported by the system cover most graph analysis operations~\cite{lee2006graphtaxonomy, srinivasan2018orko}, such as making selections, manipulating visual encodings, and performing arithmetic queries, as shown in~\autoref{tab:commands}. The system also supports chained commands that combine multiple operations in a single instruction (e.g., ``Color female nodes in red'' implicitly selects nodes and sets color). Selections can be performed on both data attributes (e.g., node or edge properties) and visual attributes (e.g., color or shape). To enable this, the system maintains a dedicated table that tracks the current visual attributes of nodes and edges, allowing subsequent commands to reference and operate on the current visual state of the graph. 
Arithmetic commands support simple graph statistics, such as computing node degree, which can be combined with aggregation operations such as mean, sum, and median to summarize graph properties.
The system provides default parameter values for certain commands to reduce user effort. For example, the command ``color nodes by gender'' automatically applies a built-in categorical color scale defined by the system unless the user specifies a different mapping.
While the command set is not intended to be comprehensive, it is sufficient to support the range of interactions needed for our research-through-design study.

\subsection{Auxiliary Views}
\paragraph{Dashboard and Session Gallery}
We designed a dashboard with buttons that allow interactions that are related to graph operations, such as creating or deleting new sessions or clearing the view (\autoref{fig:pov}-f). The dashboard also allows users to save intermediate results to the session gallery (\autoref{fig:pov}-e). The session gallery displays 2D versions of authored subgraphs saved by the user. The sessions serve as provenance~\cite{fujiwara2018provenance} and can be clicked to restore. Together, the dashboard and session gallery are designed to support users in managing their analysis provenance.
\begin{figure}[h!]
    \centering
    \includegraphics[width=0.95\columnwidth]{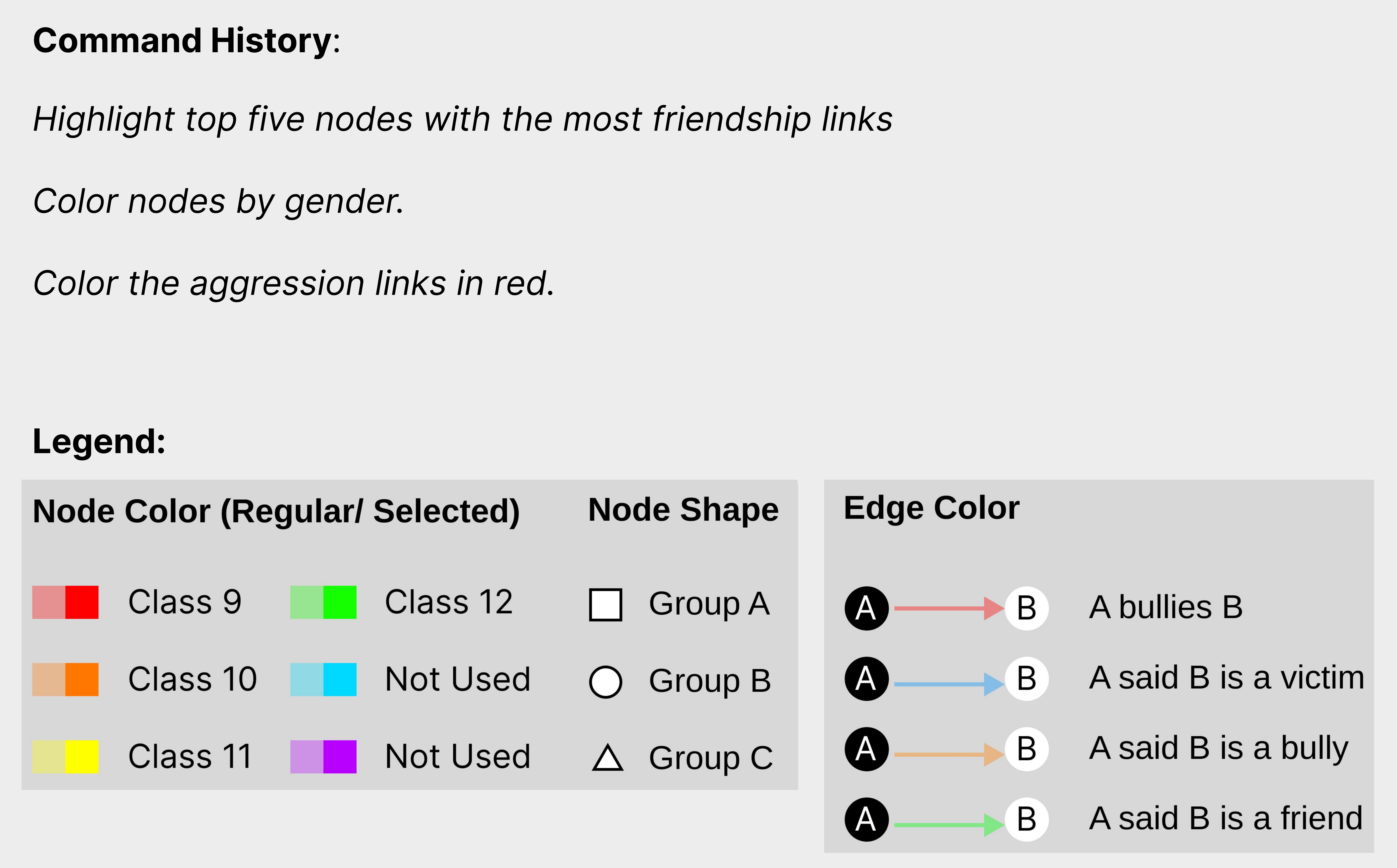}
    \caption{The text panel displays (1) command history for improved provenance, and (2) a legend showing visual encodings specified by the user through voice commands. }
    \label{fig:text_panel}
\end{figure}

\begin{figure*}[t]
    \centering
    \includegraphics[width=\textwidth]{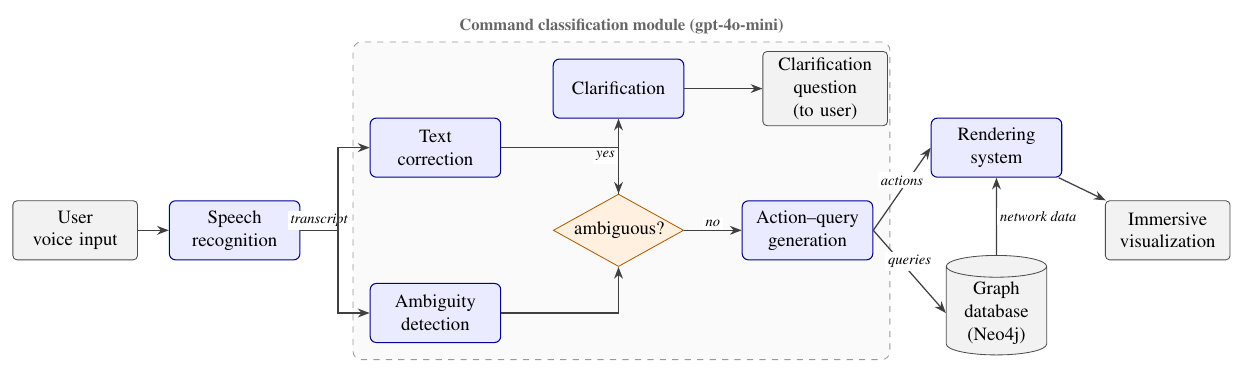}
    \caption{System architecture. The user's voice input is first transcribed into text by the speech recognition module. A command classification module then maps the utterance to rendering commands and graph queries: two stages---text correction and ambiguity detection---run concurrently on the transcript, and an unambiguous command is passed to a combined stage that generates the action sequence and corresponding graph queries in a single call. If ambiguity is detected, the system instead returns a clarification question to the user. The generated query is sent to a graph database to retrieve network data, and the rendering system combines the rendering commands with the retrieved data to render the network visualization. }
    \label{fig:pipeline}
\end{figure*}
% Along the right wall is a gallery displaying picture frames representing different sessions. 
%
% Functionally, these frames act as buttons, where pointing at a frame with the controller and clicking the controller's grip button toggles the corresponding session to be active or inactive.
%
% Additionally, each frames has a scaled-down version of its subgraph to function as a thumbnail icon for easy recognition of each session.
%
% Each frame is also labelled with the bluebluevoice query used to create them.

\paragraph{Text Panels}
% What does the chat panel do in general?
% transcription text and command visibility
The text panel displays necessary textual information on a 2D plane (\autoref{fig:text_panel}). 
First, the transcription texts from user audio are displayed in real time, accompanied by system notifications indicating the status of speech recognition and command execution.
% auto-generated legends
% We also employed shape, hue, and color saturation as complementary visual encodings to represent multiple attributes in a VR environment. 
Second, a legend shows the current encodings, including shape, hue, and color saturation, as specified by the user. 
Additionally, a command panel shows users possible commands with examples (\autoref{fig:pov}-a) to improve the discoverability of voice commands~\cite{srinivasan2019discovering_nl_commands}.
% We use discrete hues and a linear version using color scales for continuous data. Edge colors can also be encoded as different relationship types (e.g., bullying, friendship, and reported perceptions), enabling rapid identification of interaction semantics. 
% The text panel is designed to be compact and spatially stable, supporting efficient exploration of complex social networks in VR.

\subsection{System Architecture}
% Whisper -> Python built LLM agent (langgraph) -> Unity Execution
In order to support natural language interaction, we designed a voice-to-command pipeline that translates voice into graph visualization operations in real time. The overall system architecture is illustrated in ~\autoref{fig:pipeline}. 
When the user issues a voice command, the audio is transcribed into text and processed by a multi-agent pipeline that interprets the command and generates the corresponding visualization operations. These operations are then translated into database queries and executed within the VR environment.

% \subsubsection{Speech Recognition}
% Users' voice input is captured through the Oculus headset using Unity's XR Interaction Toolkit~\cite{InteractionToolkit} and transcribed through OpenAI's Whisper model~\cite{radford2022robustspeechrecognitionlargescale} in real-time. 
% Because speech recognition models may occasionally misinterpret domain-specific terminology (e.g., ``node'', ``aggression link'', or dataset attribute names), the transcription output is later processed by the preprocessing agent in the command classification module, where potential transcription errors are corrected before further analysis.

\subsubsection{Command Classification Pipeline}
A command classification pipeline with four stages, orchestrated with LangGraph~\cite{githubLangChain}, maps a transcribed utterance to (i) an ordered list of graph actions drawn from a fixed action vocabulary and (ii) one Cypher query per action. The first two stages (text correction and ambiguity detection) run in parallel to reduce latency. All stages use \texttt{gpt-4o-mini} at temperature~0. The four stages are:
\begin{enumerate}
    \item \textbf{Text correction} runs on the raw transcript. Voice commands produced by automatic speech recognition (ASR) often contain transcription errors or inconsistent wording, especially for domain-specific terminology. An LLM prompt identifies specific terms (e.g., ``node'', ``aggression link'', and dataset attribute names) and corrects likely transcription mistakes, reducing error propagation downstream.

    \item \textbf{Ambiguity detection} runs concurrently with text correction on the same transcript. A binary classifier decides whether the utterance names a concrete action and a resolvable target. Voice commands in exploratory analysis are frequently underspecified, as users may omit parameters or refer to elements implicitly (e.g., ``highlight this group'' or ``color the node'').

    \item \textbf{Clarification} is invoked conditionally. If the command is judged ambiguous, the pipeline routes to a clarification stage that returns a short follow-up question instead of executing an operation, preventing under-specified commands from propagating errors.

    \item \textbf{Action--query generation} handles unambiguous commands. A single call produces the action sequence (e.g., \texttt{selectNode}, \texttt{colorByAttribute}, \texttt{colorLink}) and the corresponding queries for the Neo4j database~\cite{neo4jNeo4jGraph}, with an empty query for actions fulfilled client-side. We retain the action sequence as an explicit intermediate representation that decouples intent interpretation from execution and eases extension to new operations and backends, but generate it together with the queries in one call to remove a sequential round-trip.
\end{enumerate}

\paragraph{Execution and Latency}
The rendering system combines the generated actions with the queried data to update the visualization. Because text correction and ambiguity detection run concurrently and action and query generation share a single call, the end-to-end model latency is not a simple accumulation over the four stages. Our technical evaluation (\autoref{sec:tech_eval}) measures a model-and-database latency of roughly 1.73\,s on average, which participants found acceptable during the user study (\autoref{sec:user_study}).

\subsection{Implementation}
The immersive system is rendered using Unity~\cite{unity} (2022.3.62f2) and its built-in XR (VR and AR) library, the XR Interaction Toolkit~\cite{InteractionToolkit} (v3.0.5).
Most of the 3D models used in the system, including the room and dashboard, were created and designed by the paper's authors using Unity's scene editor and Blender~\cite{blender}. 
%
% A few of the 3D models, like the hand controllers, are part of the XR Interaction Toolkit.
% Nodes and text surfaces was rendered using Unity's builtin shaders.
%
% For example, the nodes are spheres and the text surfaces are rectangular prisms rendered using Unity's default shaders.
%
To render the links efficiently, we implemented custom compute shaders
that generate flat rectangles representing links (always facing the user so as not to appear flat) within the GPU, to avoid rendering large amounts of 3D polygons.
%
% Instead of rendering 3D polygons, a compute shader was responsible for generating flat rectangles representing links (always facing the user as to not appear flat) within the GPU.
%
This way, the graph could be rendered in the GPU without bottlenecking the CPU, significantly improving scalability.

% Additionally, to achieve the look of the tooltip (\autoref{fig:pov}-g) appearing over other objects in the scene, even objects in front of the tooltip, the tooltip was rendered to a separate camera before being composited on top of the main camera, thereby allowing the tooltip to bypass depth sorting and avoid being occluded by other objects in the scene.
\subsection{Driving Dataset}\label{sec:dataset}
For this work, we obtained a bully-friendship network dataset from previous research in sociology~\cite{faris2014casualties, faris2020friends}. 
This dataset is representative of social networks commonly collected and analyzed by sociologists.
In this dataset, each node represents a student and each edge represents either a bullying or a friendship relationship between two students. Node (student) attributes include gender, grade level, smoker/drinker status, and GPA. 
We chose this dataset because of the real-world insights that have been uncovered by previous sociology studies, such as the association between bullying behavior and social status among students. The dataset is also used in the illustrative use case (\autoref{sec:use_case}) and the user study (\autoref{sec:user_study}). 
The system is designed to be compatible with any network dataset of a similar structure and is not limited to the specific dataset used in this study. In particular, it supports typical static networks composed of nodes and edges with associated attributes. We discuss the generality of the system further in~\autoref{sec:limitations}.

\section{Illustrative Use Case}\label{sec:use_case}

% i'll rewrite this part

% a walk through of how to use the system in a realistic scenario
% Also would be good if this case study uses a different dataset than user study
In this section, we demonstrate the system interactions on social network analysis using the dataset introduced in~\autoref{sec:dataset}.

Suppose Alice, a sociologist, is analyzing the dataset for the research question: \textit{Is bullying mostly happening within or across behavioral groups, e.g., smoking behavior?} 
To investigate this question, she issues two voice commands: ``\texttt{Highlight smokers in blue}'' and ``\texttt{Color aggression links in orange}''. These encodings allow her to visually identify the distribution of smokers and their aggression relationships with others.

\begin{figure}[t]
    \centering
    \includegraphics[width=\columnwidth]{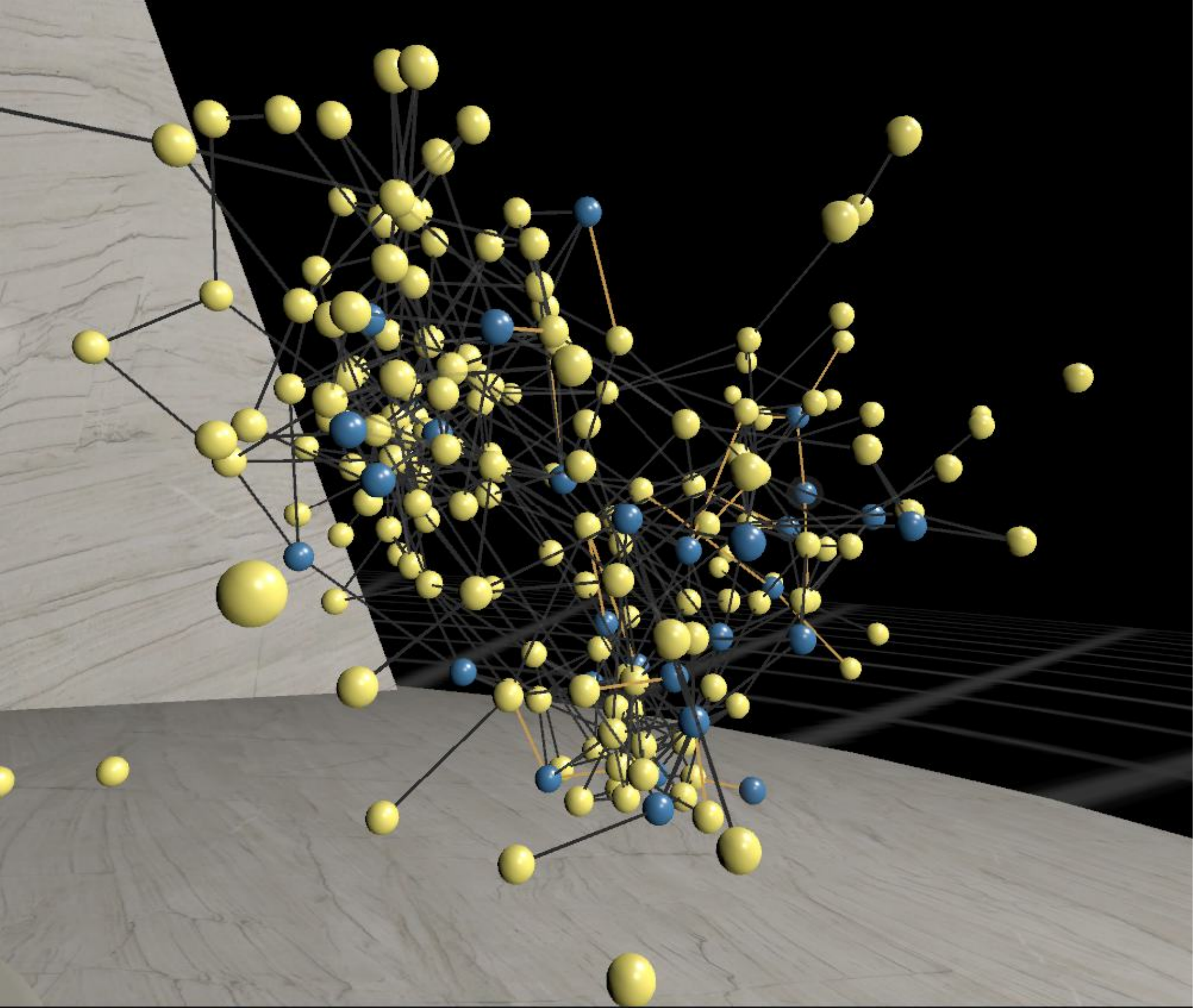}
    \caption{Visualization generated in the illustrative use case by two voice commands: ``\texttt{Highlight smokers in blue}'' and ``\texttt{Color aggression links in orange}''. Smoker nodes are highlighted in blue and aggression links are highlighted in orange, revealing bullying relationships within or across behavioral groups in the network.}
    \label{fig:smoker_case}
    \vspace*{-0.2cm}
\end{figure}

The system renders the network as shown in~\autoref{fig:smoker_case}. 
Through this visualization, Alice can observe several visual patterns.
First, regarding the distribution of smoker nodes across the network,
Alice can see that smoker nodes appear scattered throughout the network rather than forming a clearly separated cluster.
Second, smokers tend to have higher numbers of aggression links.
Third, the highlighted aggression links connect smoker nodes to both smoker and non-smoker nodes. This visual pattern suggests that bullying interactions are not limited to a single behavioral group and can occur across different behavioral categories.

This example illustrates how the system supports exploratory investigation of social network structures by translating analytical questions into visualization operations via voice commands and reveals relational patterns in an immersive environment.

\section{Technical Evaluation}\label{sec:tech_eval}
A core requirement of our voice interface is that the command pipeline correctly interprets voice input. We therefore evaluate how well the system maps a transcribed utterance to (i) an ordered list of graph actions from a fifteen-action vocabulary and (ii) one Cypher query per action. This evaluation isolates the language component: it measures the mapping from transcribed text to actions and queries, executing the generated Cypher against a live graph database, but excludes speech recognition upstream and VR rendering downstream.

\subsection{Setup}
\paragraph{Corpus}
We evaluate on a corpus of 175 labeled utterances built in three waves.
The \textbf{balanced core} (120 cases) provides roughly ten cases for each functional category of the supported voice commands, spanning node and link selection, coloring and visual encoding, movement, layout, session management, aggregate queries, multi-step composites, ASR-noise robustness, and ambiguous commands.
The \textbf{held-out adversarial expansion} (49 cases) covers the rarest actions, the decision boundaries, unseen ASR garbles, broader ambiguity, and deeper action chains, after preliminary testing on the core cases.
Finally, to identify the boundary of system performance, the \textbf{boundary probes} (6 cases) use commands whose key term names a network-analytic concept the action vocabulary cannot realize, e.g., cycle detection or pathfinding. For these cases, the correct response is to invoke clarification rather than coerce the term into an action. These cases probe whether the system recognizes the limits of its own vocabulary.
Across the first two waves, the corpus carries 186 expected action instances over 155 concrete cases plus 14 ambiguous cases.
To prevent leakage, no test cases appear in prompts, and every held-out case is verified absent from all prompt text.

\paragraph{Procedure}
Each test case is repeated three times, totaling 525 invocations on \texttt{gpt-4o-mini} at temperature 0. Correctness is scored on the first repeat, while stability and latency are aggregated over all repeats. Every generated query was executed against a local Neo4j~\cite{neo4jNeo4jGraph} instance on the bullying network with 788 nodes and 10{,}430 typed relationships. The complete run cost \$0.16.

\subsection{Results}
\autoref{tab:eval_headline} reports performance across the three corpus waves on five dimensions: \textbf{\emph{Pass}}, the case pass rate; \textbf{\emph{Clarif.}}, accuracy of the decision to clarify or act; \textbf{\emph{Action}}, action exact-sequence match; \textbf{\emph{Cypher}}, correctness of the Cypher queries; and \textbf{\emph{Stab.}}, action-sequence stability across three repeats.

On the core wave, the pipeline achieved near-perfect scores on all five dimensions, passing every case and reaching full accuracy on clarification and action prediction, with 99.0\% Cypher correctness and 99.2\% output stability. Under the held-out adversarial expansion, performance declined only slightly, with 89.8\% of cases passing, 91.1\% action exact-match accuracy, 93.6\% clarification accuracy, and 100\% Cypher correctness and stability. Since these cases were constructed to be adversarial, we regard these figures as conservative estimates of generalization to unseen phrasings. 

 On the boundary probes, the pipeline clarified only three of the six requests and coerced the remainder into fabricated actions. The reduced output stability of 83.3\% reflects this behavior, as the fabricated queries varied across repeats. These cases were designed to expose a coercion failure mode that the two primary waves under-represent, which we examine in the failure analysis below.

\begin{table}[t]
\centering
\footnotesize
\begin{tabular*}{\columnwidth}{@{\extracolsep{\fill}}l c c c c c@{}}
\toprule
\textbf{Wave} & \textbf{Pass} & \textbf{Clarif.} & \textbf{Action} & \textbf{Cypher} & \textbf{Stab.} \\
\midrule
Core (120)      & 100\% & 100\%  & 100\%  & 99.0\% & 99.2\% \\
Expansion (49)  & 89.8\% & 93.6\% & 91.1\% & 100\%  & 100\% \\
Boundary (6)    & 50\%  & 50\%   & ---    & ---    & 83.3\% \\
\bottomrule
\end{tabular*}
\caption{Pipeline performance by corpus wave, in order of increasing difficulty; see text for column definitions. Action and Cypher shape are not applicable to the boundary wave, where the correct response is to clarify rather than emit an action.}
\label{tab:eval_headline}
\end{table}

\paragraph{Latency}
\autoref{tab:eval_latency} reports per-stage latency. Note that the text-correction and ambiguity-detection stages run concurrently, so together they contribute their maximum ($M=0.66\,s$) rather than sum. Query execution adds only a small tax ($M=0.09\,s$), with a tail driven by whole-graph link selection materializing up to 10{,}430 relationships. Together the model and database average 1.73\,s, and this figure held steady as the corpus grew 41\%. We measure the pipeline latency excluding speech-to-text transcription because the transcription model (Whisper) is an interchangeable, off-the-shelf component whose latency depends on the chosen model and hardware rather than on our design.
Adding the transcription time (roughly 1\,s) observed during development and user study, user-experienced latency is around 2.73\,s. 

\begin{table}[t]
\centering
\small
\begin{tabular*}{\columnwidth}{@{\extracolsep{\fill}}l c@{}}
\toprule
\textbf{Stage} & \textbf{Latency (mean $\pm$ SD, s)} \\
\midrule
Text correction + Ambiguity detection & 0.66 $\pm$ 0.48 \\
Action--query generation & 0.98 $\pm$ 0.39 \\
Query execution (Neo4j) & 0.09 $\pm$ 0.14 \\
\textbf{Total} (model + database) & \textbf{1.73 $\pm$ 0.64} \\
\bottomrule
\end{tabular*}
\caption{Per-stage latency in seconds. The reported stages cover the command classification pipeline and database execution latency only; speech-to-text latency is excluded.}
\label{tab:eval_latency}
\end{table}

\paragraph{Output stability}
Across three repeats at temperature 0, 168/169 cases produced an identical clarification judgment and action sequence (99.4\%), and 166/169 additionally produced identical query text (98.2\%). Decision-level output is effectively deterministic, with residual variation confined to query-string surface form.

\subsection{Failure Analysis}
The technical evaluation delineates the limits of the system and yields a design lesson for voice-primary interfaces. The result shows that the system is generally reliable within its intended scope. 
Still, we found three types of errors during the evaluation. 
The first type of error occurs when an under-specified or atypically phrased command is occasionally coerced into the nearest available action.
The second type of error occurs when commands are mis-routed between execution and clarification, rather than interpreted as intended. For example, ``which student has the most friends'' is a concrete request whose answer is a set of nodes, yet it was routed to clarification instead of being executed. 

Both kinds of error are infrequent and trace to narrow gaps in the prompts or query templates rather than to the design. While such gaps can be closed by refining these components, the open-ended space of natural-language phrasings means no fixed set of rules can anticipate every input. Sustaining high accuracy therefore requires continued monitoring and iterative improvement of the system components.

The third type of error, which is more revealing, occurs when the system encounters commands that fall right around the boundary of the action vocabulary, making it hard for the system to decide if they are valid. 
For example, when a command invokes a network-analytic concept the system cannot express, it tends to fabricate a plausible but ungrounded query rather than ask for clarification. Because such queries are structurally valid and execute without error, the result can appear correct to the user even though it does not realize the requested operation. This indicates that a voice-primary, LLM-driven interface will answer confidently beyond its capabilities, underscoring the need for explicit mechanisms to recognize and surface such uncertainty. We discuss this limitation and future work further in \autoref{sec:limitations}.

% \subsection{Threats to Validity}
% Cypher is asserted to structural validity, executability, and result \emph{shape}, but not row content; 46 of 181 labelled positions remain unverifiable on the seeded graph because correct queries legitimately return zero rows there. Correctness is scored on the first of three repeats, bounded by 99.4\% decision stability. The corpus is English, single-utterance, and grounded in one schema domain, and several actions still have $\leq$7 instances; an independent corpus collected from naive users in the VR system would be a stronger generalization test. The reported latency excludes speech-to-text and rendering, which add to the end-to-end user experience beyond the $\sim$1.73\,s mean measured here. Results are specific to \texttt{gpt-4o-mini} at temperature 0; model upgrades require re-running the suite.
\section{User Study}\label{sec:user_study}
We conducted a user study to understand the user experience of our voice-based immersive network analytics system. The study is not meant to evaluate system completeness and usability, but to reveal insights into voice-based interactions through usage scenarios.
% The study focused on social network analysis, a common analysis task in sociology. 
% The study focused on participants' perceptions of voice commands as a primary interaction modality and their reflections on differences between 3D VR visualization and traditional 2D tools.
\subsection{Study Design}
We conducted a task-based evaluation complemented by qualitative interviews. Instead of a controlled usability or performance study, this approach allows us to qualitatively examine user experience on perceived usability, intuitiveness, and interaction preferences.
Participants used the social network dataset introduced in~\autoref{sec:dataset}.

\paragraph{Participants}
Ten participants were recruited through email and referrals. All participants were students in sociology (n=7) and computer science (n=3). Five participants had prior VR experience, mainly through VR games, and two of the participants had previous experience in VR-related research. This distribution provides complementary expertise in social network analysis and technical familiarity with immersive systems.

\paragraph{Task Design}
We designed three analytical tasks that explore patterns in structures and attributes within the bully-friendship network. For each task, participants were required to construct network visualizations, explore and interpret the networks, and collect answers to the tasks. The tasks are defined as:
\begin{enumerate}
    \item \textbf{Centrality and Grade Distribution.} 
    Participants examined whether socially central students were concentrated in particular grade levels and whether they were involved in bullying behavior. This task required identifying highly socially connected nodes for their attributes and aggression links.

    \item \textbf{Bullying Within or Across Behavioral Groups.} 
    Participants investigated whether bullying interactions primarily occurred within or across behavioral groups (e.g., smoker and drinker status). This task involved applying attribute-based encodings and analyzing aggression link patterns.

    \item \textbf{Isolation, Gender, and Victimization.} 
    Participants explored whether socially isolated students of a particular gender were more likely to be targeted. This required identifying nodes with fewer friendship connections but multiple incoming aggression links and examining their gender attributes.
\end{enumerate}
Tasks are designed to be open-ended, requiring participants to formulate their own voice commands to retrieve and encode relevant information. The goal was not to measure task completion time or accuracy, but to observe interaction strategies and the user experience of voice-based interaction.

\paragraph{Procedure}
Sessions were conducted in person on a Meta Quest 3 and lasted 55 minutes on average. After a background survey, participants were introduced to the system and given a hands-on tutorial on issuing voice commands (10 min). They then completed the three tasks independently in randomized order (25 min), using voice as the primary means of issuing commands while the controller remained available for node selection and dashboard operations. A closing semi-structured interview (20 min) covered their experience with voice interaction, comparisons with 2D tools such as Cytoscape~\cite{shannon2003cytoscape} and R~\cite{R}, and voice-versus-text preferences; for each question participants gave a Likert-scale rating~\cite{joshi2015likert} and explained it. Interviews were audio-recorded, transcribed, and analyzed thematically~\cite{braun2006thematic_analysis}, and participants were compensated \$20.

\subsection{Findings}
% briefly explain the analysis process
% \subsection{Themes}
\subsubsection{Positive Overall Experience}
\begin{figure}
    \centering
    \includegraphics[width=\columnwidth]{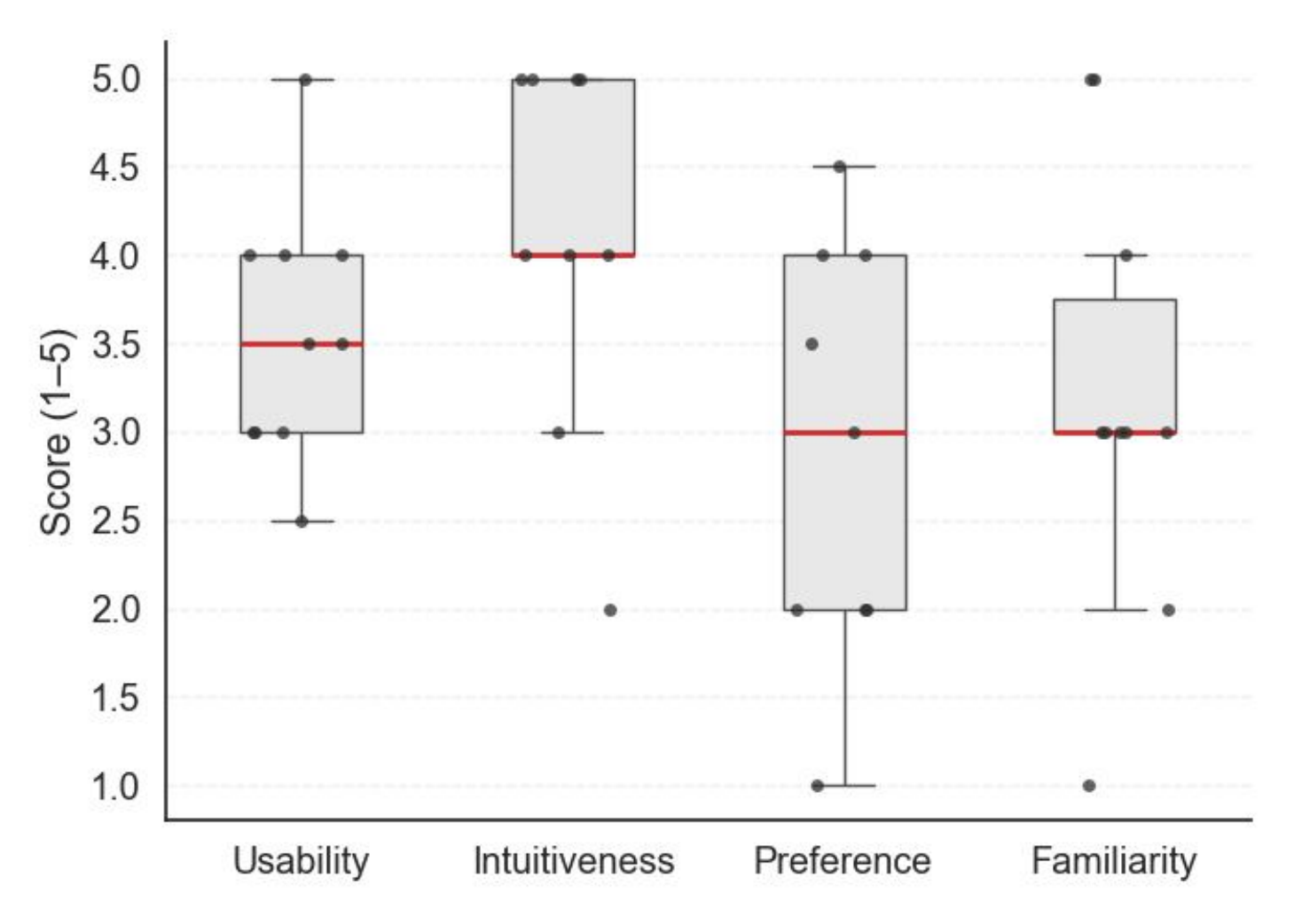}
    \caption{Distribution of participant ratings (1–5 Likert scale) for usability, intuitiveness, preference, and familiarity. Intuitiveness received the highest and most consistent ratings, while usability was generally positive with moderate variability. Preference and familiarity exhibited a greater spread, suggesting more diverse user opinions. }
    \label{fig:box_plot}
\end{figure}

Participants generally found interacting through voice and exploring the network in an immersive environment intuitive, as reflected in the questionnaire responses (\autoref{fig:box_plot}).
Intuitiveness received the highest and most consistent ratings, while usability was positive with moderate variability.
% However, one participant with no prior experience in network analysis reported that the links were sometimes confusing.
% \textit{``I sometimes don't understand the links... the regression links between the nodes (P10).''}

All participants agreed that voice commands were easy to use, and several noted that voice lowers the barrier for users unfamiliar with analytical software: \textit{``Voice is definitely way more approachable than trying to plot the graph in R using \dots the SNA package (P4).''} A few gave lower usability scores because precise node selection with the controller was difficult, which we examine next.

% During the interviews, six participants mentioned that interacting with nodes using controllers was more difficult than issuing voice commands, particularly when the network becomes larger and denser. Participants without prior VR experience (N=5) reported that they needed some time to familiarize with the headset and controllers, which influenced their usability and preference ratings. 
% As they explained: \textit{``usability maybe like three just because I haven't really used it before, so it's a little bit hard to get used to the headset and using the controller (P8).''}

% In contrast, participants consistently reported that the voice interaction was intuitive. 
% All participants agreed that voice commands were easy to use when interacting with the system. One noted that voice interaction could be particularly helpful for users who are unfamiliar with analytical software: \textit{``If you're a sociologist, you might not have ever used any kind of software before. Um, so being able to just like talk to it is good. (P4)''}
% And I think the network graphs are such a, like anyone who's starting to use social network analysis is always so interested in the graphs first (P4)."}

Participants found the response latency acceptable, likening it to ``a common process when using some other tools (P10),'' though P6 noted that the system occasionally misread a less explicit command.
% controller precision, familiarity with VR, and the accuracy of voice command recognition, all contributed to the usability and intuitiveness ratings reported in the study.

\subsubsection{Voice Interaction vs. Controller Interaction}
All participants preferred issuing voice commands to typing, including those new to VR. Six found selecting nodes with the handheld controllers more difficult than speaking, especially as the network grew denser, and the five without prior VR experience needed time to adjust to the headset and controllers. In contrast, they described voice as reducing motor coordination and letting them express intent directly; as one first-time VR user put it, ``it didn't really take much for me to do anything besides speaking alone (P9).'' These observations contrast the physical demands of precise controller interaction with the simplicity of voice, suggesting voice can serve as an accessible primary modality in VR for novices.
% participants frequently described voice interaction as intuitive and accessible, particularly for users without advanced technical knowledge or prior VR familiarity. Rather than requiring precise hand control or interface manipulation, 

% For participants with limited VR experience, voice interaction appeared to lower the entry barrier by minimizing interaction complexity. 

% , by reducing both motor demands and perceived technical difficulty.

\subsubsection{Preference on Voice Input vs. Text Input}
Participants' preferences between voice and text were shaped by their everyday communication habits. Several felt that text forces them to compress and structure a message, whereas voice allows more natural expression: \textit{``texting \dots forces me to say less of what I want. I have to summarize a lot \dots (P2).''} Because commands here are directed at a machine rather than a person, participants perceived less pressure to refine their phrasing, suggesting that voice can lower the cognitive effort of formulating commands.

\subsubsection{Social Considerations for Voice Input Adoption}
Acceptance of voice input was also shaped by social factors often overlooked in system design. Participants raised privacy concerns about speaking commands in public, yet valued the emotional expressiveness of voice over text (P9, P6), and noted that native-language fluency can make speaking easier than typing for multilingual users (P5). These factors may shape whether users adopt voice interaction.

\subsubsection{Improved Spatial Understanding and Engagement}
% We compared participants’ user experiences using the traditional 2D interface and our 3D VR system. Overall, participants reported differences in spatial understanding, engagement, and interaction effort.

% \textit{``I think that it's a lot easier to visualize like in a 3D way because there's like overlap and it's hard to visualize in a 2D like plane.(P2)."} 
Compared to 2D interfaces, participants reported better spatial understanding and engagement in the immersive visualization. Contrary to the known limitations of 3D on 2D displays~\cite{munzner2014visualization_analysis_design}, several found that the immersive 3D view reduced the node and link overlap that is hard to disentangle in 2D and let them walk around the structure: \textit{``with 3D \dots you can see the whole thing and you can walk around it. With 2D you can't (P8).''} Real-time voice interaction further heightened engagement, as issuing a command and immediately seeing the result made the analysis feel more interactive and immersive (P6). Together, these suggest that immersive visualization with voice interaction supports both spatial reasoning and a more engaging analytical experience than conventional 2D tools.

\section{Discussion}
% Discuss implications from user study for methods to improve usability
The user study shows the positive usability of our system. In this section, we discuss the role of usability in improving VR adoption, and the benefits of voice-based interactions for improving interaction fluidity and simplifying operations while supporting complex, multi-parameter authoring tasks.
\subsection{Usability limits adoption of immersive visualization}
Previous work has highlighted the benefits of immersive visualization, including enhanced spatial reasoning, depth perception, and engagement~\cite{kraus2021value_of_immersive_vis}. However, in practice, immersive visualization tools continue to face usability challenges, which contribute to low adoption~\cite{Ens2021grand_challenges}. 
While controller-based interactions are familiar to many VR users, our study found that voice-based interactions provided a substantially better user experience for network analysis. This is because users are relieved from learning and remembering the positions of embedded UIs while also maintaining precise and stable controller movements. As a result, voice input enables a wide range of network analysis operations without introducing excessive menus or buttons, significantly improving usability. These findings suggest that simplifying interaction through voice input can help overcome usability barriers and promote wider adoption of immersive visualization. This complements prior usability-driven efforts in immersive analytics that revisit conventional input, such as bringing mouse and keyboard into the headset to recover precision and familiarity~\cite{grubert2020back_to_the_future}. Voice offers an alternative route to the same goal: it preserves precise controller movement for direct manipulation while removing the dependence on conventional input.

% We focus on simplifying interaction while supporting more complex operations
\subsection{Voice-based interactions improve fluidity}
Fluidity in visualization refers to interaction designs that help users stay in the ``flow'', i.e., a mental state of total immersion in an activity with high focus, involvement, and rewarding outcomes~\cite{csikszentmihalyi1990flow, elmqvist2011fluid_interaction}.
While immersive visualization appears to be promising in achieving fluidity, the usability issues have remained a challenge. 
Findings in our user study show the potential of voice-based interactions in achieving fluidity. Voice-based interactions allow users to perform interactions with natural expressions, reducing the cognitive demand of organizing well-structured messages. This increases spatial understanding, engagement, and immersion. Minimizing controller interactions also prevents ``creative chaos'' as seen in previous research~\cite{derksen2025creative_chaos}.
Although the latency and recognition accuracy of voice commands limit fluidity currently, we anticipate that voice-based technologies will continue to improve in the future.
This observation also speaks to a central question in immersive analytics: whether immersion can be shaped by non-visual modalities~\cite{buschel2018interaction_for_ia, skarbez2019ia_research_agenda}. Whereas that question has largely concerned output, our findings suggest that input matters as well: voice-based interaction increases fluidity and, in turn, the sense of immersion~\cite{witmer1998presence}. This broadens the envisioned design space, in which immersion is shaped not only by how data is presented but also by how the analyst acts upon it.
% While the latency and recognition accuracy of voice commands remain the bottleneck for further improving fluidity, we expect voice-based technology will readily improve in the foreseeable future.

% Achieving fluid interaction in immersive environment remains a challenge, as we see limitations of voice interactions

\subsection{Voice commands may reduce physical exploration}
Since our system design emphasized voice-based interaction, participants often expected the system to bring the network subset of interest closer to them for examination and manipulation. As a result, most participants remained largely stationary during the tasks, not fully realizing that they could move around the visualization space or physically navigate into the network structure itself. This behavior suggests that prioritizing voice commands may implicitly discourage users from engaging in embodied exploration~\cite{batch2020there_is_no_spoon, cordeil2017imaxes} as an interaction strategy in immersive environments.

Importantly, this emphasis on voice and the resulting behavior does not affect our main findings, but it highlights a limitation of the current design. Immersive environments likely benefit from hybrid approaches that combine voice with embodied navigation and controller-based manipulation: voice efficiently supports high-level analytical commands, while physical navigation and controller input facilitate spatial exploration and precise object manipulation~\cite{drogemuller2018navigation_3d_graph_vr}, letting users switch between conversational control and embodied interaction as task demands change. Exploring such hybrid designs may further enhance analytical flexibility and engagement, which we leave for future work.

\subsection{Limitations and Future Work}\label{sec:limitations}
\paragraph{Participants and Evaluation Scope}
Our study recruited ten student participants, a small sample drawn largely from a non-expert population. Their limited prior exposure to VR, network visualization, and voice interaction may also introduce a novelty effect into the interview responses. Since we did not collect the precise time of voice versus controller usage during tasks, our observations on modality preference rest on participants' self-reports and experimenter observation. As a research-through-design study~\cite{zimmerman2007research_through_design} of an underexplored interaction paradigm, our aim is to characterize the design space and surface phenomena that warrant future investigation, rather than to establish performance benefits. 
Accordingly, the technical evaluation (\autoref{sec:tech_eval}) quantifies the pipeline's behavior, while the user study characterizes experience qualitatively. A controlled, quantitative study of voice-primary interaction, ideally with experienced analysts as intended users, is an important next step that our findings are intended to inform.

\paragraph{Handling Ambiguity}
Our study assumes that LLM-based speech recognition technologies can tolerate a reasonable level of ambiguity without substantially affecting usability. However, certain forms of ambiguity may still require dedicated interaction designs to resolve effectively~\cite{gao2015datatone}. While we did not explore such mechanisms in this work, our ambiguity detection agent helps prevent unclear commands from propagating errors through the system. In our study, this safeguard was sufficient to maintain a stable user experience.

\paragraph{Discoverability}
Natural language interfaces, including voice commands, inherently face discoverability challenges~\cite{srinivasan2019discovering_nl_commands}, as users may be uncertain about which commands the system supports. In our study, we mitigated this issue through onboarding tutorials and an ``Example Command'' panel. However, these solutions remain external aids rather than integrated interaction mechanisms. Designing more natural and context-aware approaches for improving command discoverability remains an important future direction.

\paragraph{Advanced Network Analysis}
Our current system focuses on supporting common network exploration tasks and does not address more advanced analytical operations, such as clustering, or more complex network types, including extreme-scale, geospatial, dynamic, or hierarchical networks~\cite{sorger2019large_dynamic_networks_vr}. Enabling voice interaction for these scenarios would introduce additional technical and interaction challenges and may require new interaction designs and usability considerations, which we leave for future work.
Our technical evaluation makes this boundary concrete: commands naming unsupported analytic concepts (e.g., cycle motifs or brokerage roles) are correctly clarified only half the time, with the remainder coerced into unrelated operations (\autoref{sec:tech_eval}). Bridging this gap would require a dedicated graph-analytics layer (cycle, path, and centrality operators) beneath the voice interface. Since our study concerns the usability and fluidity of voice-primary interaction, we expect this limitation to have little bearing on our main conclusions.

\section{Conclusion}
In this work, we present a research-through-design study on a voice-based interaction system for immersive visual exploration and analysis of network data. 
Our work highlights several design implications for future voice-based analytical systems. First, voice commands are particularly effective for improving usability and fluidity for high-level analytical operations. 
Second, designers should account for social factors that shape users' preferences to encourage adoption of voice interaction.
Third, systems should provide mechanisms to mitigate ambiguity and improve command discoverability. Together with existing guidelines for voice-based interfaces, these findings suggest that effective immersive analytics systems should balance conversational interaction with visual and embodied interactions, enabling users to fluidly navigate analytical tasks.

% \section*{Acknowledgments}
% This project is supported in part by the ABC agency with grant number NNN.
% The authors appreciate the dataset and feedback on visual analytics tasks provided by collaborators from Sociology.
%% if specified like this the section will be committed in review mode
\acknowledgments{
This research is supported in part by the U.S. National Science Foundation via grant No. IIS-2427770 and by the Argonne National Laboratory via contract No. 6F-60084.
}

\bibliographystyle{abbrv-doi}

\bibliography{references}

\clearpage
\appendix
\section{Agent Prompts}\label{app:prompts}
For replicability, we report the system prompts that define each stage of the command classification pipeline (\autoref{sec:tech_eval}). All stages run on \texttt{gpt-4o-mini} at temperature~0; the action--query generator additionally constrains decoding to a JSON object. The prompts are reproduced verbatim below; the complete source, including the live examples, is available in the released repository.

\lstset{
  basicstyle=\ttfamily\scriptsize,
  breaklines=true,
  breakindent=0pt,
  columns=fullflexible,
  frame=single,
  framesep=4pt,
  xleftmargin=4pt,
  showstringspaces=false,
  keepspaces=true,
}

\subsection{Text Correction}
\begin{lstlisting}
You are correcting ASR (voice recognition) errors for graph visualization commands.

Rules:
- Fix ONLY misheard words (e.g., note->node, blew->blue, caller->color, great->grade, sacks->sex, gee pee ay->GPA, GP->GPA, geepay->GPA)
- NEVER convert color names to hex codes. Keep "red" as "red", "blue" as "blue", etc.
- NEVER change the word "color" - it is a verb meaning "to paint/color".
- Layout names: the app has exactly three layouts - "floor", "cluster", and "spherical". When a word sounds like a mishearing of one of these layout names, replace it with the canonical name. A single layout name may also be transcribed as a run of two or three smaller words (the way "gee pee ay" is a transcription of "GPA"); rejoin and correct such runs to the canonical layout name.
- DO NOT add or remove words. Only fix misheard ones. (Rejoining a multi-word mishearing of one term, as above, counts as fixing, not removing.)

Examples:
- "color nodes by grade" -> "color nodes by grade"
- "colour the notes blew" -> "color the nodes blue"
- "select the notes with blew caller" -> "select the nodes with blue color"
- "highlight top 3 notes by friendship" -> "highlight top 3 nodes by friendship"

Return ONLY the corrected text. Nothing else.
\end{lstlisting}

\subsection{Ambiguity Detection}
\begin{lstlisting}
You decide whether a voice command for a graph-visualization tool is too vague to act on.

Reply ONLY one word:
- "yes" -> ambiguous: the intended ACTION or TARGET cannot be determined
- "no"  -> clear: a concrete action is identifiable

Reply "yes" when the command names no concrete operation, or no resolvable target, e.g.:
  "do the thing", "change it", "make it look better", "fix this", "select the one over there", "do that again"

Reply "no" when a concrete action is identifiable (select / color / shape / size / move / layout / deselect / reset / save / delete) - even if it refers to the current selection with words like "them", "it", "these", or "the selected", e.g.:
  "color nodes by grade", "shape nodes by sex", "make the selected nodes blue", "deselect all", "save session as top athletes"

A recognizable verb makes a command clear even if its object is the current selection or its target is loosely worded:
- movement/layout: "bring them closer to me", "project them onto the floor"  -> clear
- reset synonyms: "reset", "start over"  -> clear
- counts/aggregates: "how many nodes are selected", "what's the average GPA"  -> clear (arithmetic)

Voice-misheard words (colors, attribute names, etc.) are NOT ambiguous - they are corrected downstream. Default to "no" unless the command is genuinely vague.
\end{lstlisting}

\subsection{Clarification}
\begin{lstlisting}
User request may be unclear.

Ask ONE short clarification question.
\end{lstlisting}

\subsection{Action--Query Generation}
The action--query generator emits a JSON object with two index-aligned arrays, \texttt{actions} (a list of \texttt{[actionName, param]} pairs over the fifteen-action vocabulary) and \texttt{queries} (one Cypher string per action, empty for client-side actions).
\begin{lstlisting}
You are a graph visualization assistant. Given a user command, return a JSON object with exactly two keys:
- "actions": list of [actionName, param] pairs
- "queries": list of Cypher query strings, one per action ("" for actions that need no database query)
The arrays must be the same length and index-aligned.

ACTIONS
Allowed action names:
  selectNode, selectLink, colorNode, colorLink, colorByAttribute, colorByGPA, shapeByAttribute, sizeNode, move, layout, deselect, arithmetic, reset, saveSession, deleteSession

-- COLOR NAME -> HEX --
When generating colorNode or colorLink, convert color names to hex (red->#FF0000, blue->#0000FF, ...). Any hex color may also be passed directly.

-- GPA COLORING (highest priority) --
ANY mention of "GPA" in a coloring context -> ALWAYS use colorByGPA, never colorByAttribute.
  "color nodes by GPA" -> [["colorByGPA","gpa"]]

-- "color BY attribute" vs "color IN a color" --
  "color nodes by grade"       -> [["colorByAttribute","grade"]]   (categorical)
  "color selected nodes red"   -> [["colorNode","#FF0000"]]        (single color)
  "color aggression links red" -> [["selectLink","aggression"],["colorLink","#FF0000"]]

-- LINK SELECTION --
selectLink param = link type name ("aggression","friendship") or "all". Append ":selected" to scope to currently selected nodes.

-- NODE SELECTION --
"top N nodes by metric" -> always pair selectNode + colorNode. This pairing is MANDATORY for every top-N / superlative metric selection, whatever the verb. If the ranking metric is gpa, the paired color action is colorByGPA instead. "selected nodes" / "highlighted nodes" = use current selection, do NOT add new selectNode. Conditions can be COMPOUND: join attribute tests with AND / OR inside the ONE selectNode param. A request with multiple conditions is still ONE selectNode action.

-- OTHER --
  "shape nodes by <attr>"         -> [["shapeByAttribute","<attr>"]]
  "size nodes by <attr>"          -> [["sizeNode","<attr>:all"]]

-- deselect --
Clears the CURRENT selection (client-side). It does NOT change colors - that is reset. Emit deselect ONLY when the user explicitly asks to clear / deselect / unselect. NEVER add deselect to a color / move / size / arithmetic command that acts on "the selected" items.

-- layout vs move --
layout = arrange the nodes into one of the named layouts ("floor", "cluster", "spherical"). move = translate the CURRENT selection through space relative to the user; it takes no parameter. Movement verb + named arrangement/surface -> layout. Movement verb + a direction/place relative to the user -> move.

-- arithmetic (questions about quantities) --
ANY question asking for a quantity or statistic (count, total, average, min, max, sum) MUST produce exactly ONE ["arithmetic",""] action paired with an aggregate Cypher query. arithmetic is ONLY for requests whose ANSWER IS A NUMBER. A request to find/select the node(s) with the highest/lowest value of a metric asks for NODES, not a number - that is a top-N/superlative selectNode.

CYPHER QUERIES (index-aligned with actions)
Use "" for: colorNode, colorLink, colorByGPA, move, layout, deselect, reset, saveSession, deleteSession.
  selectNode simple:  ["selectNode","n.sex = 'female'"]  -> MATCH (n:Node) WHERE n.sex = 'female' RETURN n
  selectNode top-N:   ["selectNode","top_3_friendship_degree"] -> MATCH (n:Node)-[r:POINTS_TO]-(m) WHERE r.type='friendship' WITH n,COUNT(r) AS d ORDER BY d DESC LIMIT 3 RETURN n
  selectLink by type: ["selectLink","aggression"] -> MATCH (n:Node)-[r:POINTS_TO]-(m) WHERE r.type='aggression' RETURN r
  colorByAttribute:   ["colorByAttribute","grade"] -> MATCH (n:Node) RETURN DISTINCT n.grade AS value ORDER BY value
  sizeNode:           ["sizeNode","degree:all"] -> MATCH (n:Node) RETURN min(n.degree) AS minValue, max(n.degree) AS maxValue
  arithmetic:         count -> MATCH (n:Node) WHERE n.selected=true RETURN count(n); average -> ... RETURN avg(n.<attr>); etc. Questions about the whole graph drop the WHERE clause.

GLOBAL RULES
1. Emit ONLY the actions the user explicitly requested. NEVER append an extra action the user did not ask for (exceptions: pairings this prompt itself mandates).
2. If the input is an actionable command or question, "actions" must be NON-EMPTY.

Return ONLY a JSON object. No markdown. No explanation.
Example: {"actions":[["selectNode","n.grade=9"],["colorNode","#FF0000"]],"queries":["MATCH (n:Node) WHERE n.grade=9 RETURN n",""]}
\end{lstlisting}

\end{document}